\documentclass[aps,prl,twocolumn,groupedaddress,draft]{revtex4}
\usepackage{graphicx,epsf}
\newif\ifdraft
\drafttrue
\newif\ifpreprint
\preprinttrue

\def\fig#1{fig.~{\ref{#1}}}

\def\spa#1.#2{\left\langle#1\,#2\right\rangle}
\def\spb#1.#2{\left[#1\,#2\right]}
\def\tree{{\rm tree}}

\def\eqn#1{eq.~(\ref{#1})}

\def\NeqFour{{\cal N}=4}

\def\NeqEight{{\cal N}=8}

\def\be{\begin{equation}}
\def\ee{\end{equation}}
\def\bea{\begin{eqnarray}}
\def\eea{\end{eqnarray}}
\def\ba{\begin{eqnarray}}
\def\ea{\end{eqnarray}}

\def\eps{\epsilon}

\def\e{\epsilon}

\def\nn{\nonumber}

\def\oneloop{{\rm 1 \hbox{-}loop}}

\def\Ord{{\cal O}}

\newbox\charbox
\newbox\slabox
\def\s#1{{      % Feynman slash
        \setbox\charbox=\hbox{$#1$}
        \setbox\slabox=\hbox{$/$}
        \dimen\charbox=\ht\slabox
        \advance\dimen\charbox by -\dp\slabox
        \advance\dimen\charbox by -\ht\charbox
        \advance\dimen\charbox by \dp\charbox
        \divide\dimen\charbox by 2
        \raise-\dimen\charbox\hbox to \wd\charbox{\hss/\hss}
        \llap{$#1$} }}

\def\sandpm#1.#2.#3{%
  \left\langle\smash{#1}{\vphantom1}^{+}\right|{#2}%
  \left|\smash{#3}{\vphantom1}^{-}\right\rangle}
\def\spa#1.#2{\left\langle#1\,#2\right\rangle}
\def\spb#1.#2{\left[#1\,#2\right]}
\def\spab#1.#2.#3{\langle\mskip-1mu{#1}^- 
                  | #2 | {#3}^-\mskip-1mu\rangle}
\def\spba#1.#2.#3{\langle\mskip-1mu{#1}^+ 
                  | #2 | {#3}^+\mskip-1mu\rangle}
\def\spbb#1.#2.#3{\langle\mskip-1mu{#1}^+ 
                  | #2 | {#3}^-\mskip-1mu\rangle}
\def\lor#1.#2{\left(#1\,#2\right)}

\begin{document}

\ifpreprint
\noindent
UCLA/06/TEP/30\\
$\null$ \hfill SLAC-PUB-12187
\fi

\title{Is $\NeqEight$ Supergravity Ultraviolet Finite?}

\author{Z.~Bern${}^a$, L.~J.~Dixon${}^{b}$, R.~Roiban${}^c$}
\affiliation{
${}^a$Department of Physics and Astronomy,
UCLA, Los Angeles, CA 90095-1547, USA  \\
${}^b$Stanford Linear Accelerator Center,
Stanford University, Stanford, CA 94309, USA\\
${}^e$Department of Physics, Pennsylvania State University,
University Park, PA 16802, USA\\
}

\begin{abstract}
Conventional wisdom holds that no four-dimensional gravity field
theory can be ultraviolet finite. This understanding is based mainly
on power counting.  Recent studies confirm that one-loop $\NeqEight$
supergravity amplitudes satisfy the so-called ``no-triangle
hypothesis'', which states that triangle and bubble integrals cancel
from these amplitudes. A consequence of this hypothesis is that for
any number of external legs, at one loop $\NeqEight$ supergravity and
$\NeqFour$ super-Yang-Mills have identical superficial degrees of ultraviolet
behavior in $D$ dimensions.  We describe how the unitarity method
allows us to promote these one-loop cancellations to higher loops,
suggesting that previous power counts were too conservative.  We
discuss higher-loop evidence suggesting that $\NeqEight$ supergravity
has the same degree of divergence as $\NeqFour$ super-Yang-Mills
theory and is ultraviolet finite in four dimensions.  We comment on
calculations needed to reinforce this proposal, 
which are feasible using the unitarity method.
\end{abstract}

%\pacs{11.15.Bt, 11.25.Db, 11.25.Tq, 11.55.Bq, 12.38.Bx \hspace{1cm}}

\maketitle

${}$
\vspace{-10truemm}

\section{Introduction}

Conventional wisdom holds that it is impossible to construct a finite
field theory of quantum gravity.  Indeed, all widely accepted studies
to date have concluded that all known gravity field theories are
ultraviolet divergent and
non-renormalizable~\cite{tHooftVeltmanAnnPoin, tHooftGrav,
SusyGravity, HoweStelleReview}.  If one were able to find a finite
four-dimensional quantum field theory of gravity, it would have
profound implications.  In particular, finiteness would seem to imply that
there should be an additional symmetry hidden in the theory.

Although power counting arguments indicate that all known gravity field
theories are non-renormalizable, there are very few explicit
calculations establishing their divergence properties.  For pure
gravity, a field redefinition removes the potential on-shell one-loop
divergence~\cite{tHooftVeltmanAnnPoin, tHooftGrav}, but the
calculation of Goroff and Sagnotti~\cite{GoroffSagnotti}, confirmed
by van de Ven~\cite{vandeVen}, explicitly shows that pure Einstein
gravity has an ultraviolet divergence at two loops.  If generic matter
fields are added~\cite{tHooftVeltmanAnnPoin, tHooftGrav} a divergence appears
already at one loop.  If the matter is added so as to make the theory
supersymmetric, the divergences are in general delayed until at least
three loops (see e.g. refs.~\cite{SusyGravity,HoweStelleReview}).
However, no complete calculations have been performed to confirm that
the coefficients of the potential divergences in supersymmetric
theories are actually non-vanishing.

One approach to dealing with the calculational
difficulties~\cite{BDDPR, OneloopMHVGravity, DunbarNorridge,
InheritedTwistor, NoTriangle} makes use of the unitarity
method~\cite{UnitarityMethod,DDimUnitarity, GeneralizedUnitarity,
TwoLoopSplit, BCFUnitarity}, as well as the Kawai, Lewellen and Tye
(KLT) relations between open- and closed-string tree-level
amplitudes~\cite{KLT}.  In the low-energy limit the KLT relations
express gravity tree amplitudes in terms of gauge theory tree
amplitudes~\cite{BGK}. Combining the KLT representation with the
unitarity method, which builds loop amplitudes from tree amplitudes,
massless gravity scattering amplitudes -- including their ultraviolet
divergences -- are fully determined to any loop order starting from
gauge theory tree amplitudes.  In particular, for the case of
$\NeqEight$ supergravity, the entire perturbative expansion can be
built from $\NeqFour$ super-Yang-Mills tree amplitudes~\cite{BDDPR}.
It is rather striking that one can obtain all the amplitudes of
$\NeqEight$ supergravity from the tree amplitudes of an
ultraviolet-finite conformal field theory.

The KLT relations between gauge and gravity amplitudes are especially
useful for addressing the question of the ultraviolet divergences of
gravity theories because, from a technical viewpoint, perturbative
computations in gauge theories are much simpler than in gravity
theories.  With the unitarity method, these relations are promoted to
relations on the unitarity cuts.  This strategy has already been 
used~\cite{BDDPR} to argue that the first potential divergence in
$\NeqEight$ supergravity would occur at five loops, instead of the
three loops previously predicted using superspace power counting
arguments~\cite{HoweStelleReview}.  Using harmonic superspace, Howe
and Stelle have confirmed this result~\cite{HoweStelleNew}.  Very
interestingly, they also speculate that the potential divergences may
be delayed an additional loop order.

In this note we reexamine the power counting of ref.~\cite{BDDPR} for
$\NeqEight$ supergravity. We demonstrate that there are additional
unexpected cancellations beyond those identified in that paper.  Our
analysis of the amplitudes is based on unitarity cuts which slice through
three or more lines representing particles, 
instead of the iterated two-particle cuts 
focused on in ref.~\cite{BDDPR}. It suggests that $\NeqEight$ supergravity
may have the same ultraviolet behavior as $\NeqFour$ super-Yang-Mills
theory, {\it i.e.} that it is finite in four dimensions.  We will also
outline calculations, feasible using the unitarity method, 
that should be done to shed further light on this issue.

Our motivation for carrying out this reexamination stems from the
recent realization that in the $\NeqEight$ theory unexpected one-loop
cancellations first observed for the special class of maximally
helicity violating (MHV) amplitudes~\cite{OneloopMHVGravity}, in fact
hold more generally~\cite{InheritedTwistor,SixPtIRGravity,
NoTriangle}. The potential consequences of these cancellations for
the ultraviolet divergences at higher loops were noted in the latter
references. See also ref.~\cite{KITPTalk}.

It has been known for a while that the one-loop MHV amplitudes
with four, five and six external gravitons are composed entirely of scalar box
integrals~\cite{GSB,OneloopMHVGravity}, lacking all triangle and
bubble integrals.  Although there are no complete calculations 
for more than six external legs, factorization arguments make a 
strong case that the same property holds for MHV amplitudes with an
arbitrary number of external legs.  Factorization puts
rather strong constraints on amplitudes and has even been used to
determine their explicit form in some cases (see, for example,
refs.~\cite{FactorizationExamples, OneloopMHVGravity}).  Because the
scalar box integral functions are the same ones that appear in the
corresponding $\NeqFour$ super-Yang-Mills amplitudes, the
$D$-dimensional ultraviolet behavior is identical: The amplitudes
all begin to diverge at $D=8$.

As the number of external legs increases, one might have expected 
the ultraviolet properties of supergravity to become relatively worse
compared to super-Yang-Mills theory, as a consequence of the 
two-derivative couplings of gravity.
From recent explicit calculations of six and seven graviton
amplitudes~\cite{InheritedTwistor, SixPtIRGravity, NoTriangle},
it is now clear that the cancellations which prevent triangle or
bubble integrals from appearing extend beyond the MHV case.  This
property has been referred to as the ``no-triangle hypothesis''.  This
hypothesis puts an upper bound on the number of loop momenta
that can appear in the numerator of any one-loop integral.  Under integral
reductions~\cite{PV} any power of loop momentum $l$ appearing in the
numerator can be used to reduce an $m$-gon integral to an $(m-1)$-gon 
integral.  The $(m-1)$-gon integral has one less power of $l$ in its
numerator.  Inverse propagators (such as $l^2$) present in the 
numerator are special: they can also be used to reduce $m$-gon 
integrals to $(m-1)$-gon integrals with {\it two} less powers of $l$.

For example, consider a pentagon integral with
a factor of $2 \, l \cdot k_1$ in the numerator, denoted by $I_5[2\, l
\cdot k_1]$. If $l^2$ and $(l-k_1)^2$ are two propagators in the
pentagon integral and $k_1$ is the momentum of an external line, we
can rewrite the numerator factor as a difference of two inverse
propagators,
\begin{equation}
2\, l \cdot k_1 = l^2 - (l - k_1)^2 \,.
\end{equation}
This equation immediately reduces the linear pentagon integral to a difference 
of two scalar box integrals,
\begin{equation}
I_5[2\, l \cdot k_1]  = I_4^{(1)} - I_4^{(2)}\,, 
\end{equation}
where  $I_4^{(1)}$ and  $I_4^{(2)}$ are the box integrals
obtained from the pentagon by removing the $l^2$ and $(l-k_1)^2$
propagators, respectively. 
More generally, integral reductions bound the maximum power of 
loop momenta in the numerator, in order that
triangle integrals not appear in the final result.  For a pentagon
integral the maximum is one inverse propagator, or one generic power
of $l$; for a hexagon it is two inverse propagators,
or two generic powers of $l$; and so forth.

For six-graviton non-MHV amplitudes, the computations of
ref.~\cite{NoTriangle} constitute a proof of the no-triangle
hypothesis.  For seven gravitons they demonstrate that the box integrals
correctly account for infrared divergences, making it unlikely for
infrared-divergent triangle integrals to appear.  For larger numbers
of external gravitons, the same factorization arguments used for
MHV amplitudes~\cite{OneloopMHVGravity,InheritedTwistor} make a 
strong case that the required cancellations continue to hold for 
all graviton helicity configurations.  Supersymmetry relations suggest
that the no-triangle hypothesis should be extendable from graviton
amplitudes to amplitudes for all possible external states,
because all $\NeqEight$ supergravity states belong to the same 
supermultiplet. 
The additional cancellations implied by the no-triangle 
hypothesis are rather non-trivial.  For example, one-loop 
integrands constructed using the KLT representation of
tree amplitudes naively violate the no-triangle
bound~\cite{OneloopMHVGravity,InheritedTwistor}.
In the rest of this letter we will assume that the 
no-triangle hypothesis holds for all one-loop $\NeqEight$ amplitudes and 
examine the consequences for higher loops.

Will these unexpected one-loop cancellations continue to higher
orders? Using conventional Feynman diagram techniques, it is
not at all clear how to extract from these one-loop on-shell cancellations
useful higher-loop statements, given that the formalism is inherently
off shell.  The unitarity method~\cite{UnitarityMethod, DDimUnitarity,
GeneralizedUnitarity, TwoLoopSplit, BCFUnitarity}, however, provides a
means for doing so. Because of the direct way that lower-loop on-shell
amplitudes are used to construct the higher-loop ones, it is clear that 
the one-loop cancellations will continue to be found
in higher-loop amplitudes. The main question then is whether the
cancellations are sufficient to imply finiteness of the theory to all
loop orders.

Besides the implications of the no-triangle hypothesis for higher loops
via unitarity, there are a number of other clues pointing to a better
than expected ultraviolet behavior.  One interesting clue comes from
the fact that the only complete $\NeqEight$ calculation at two loops -- the
four-graviton amplitude -- has exactly the same power
counting as the corresponding $\NeqFour$ super-Yang-Mills
amplitude~\cite{BDDPR}.  This example clearly shows
that a common degree of finiteness between the two theories
in $D$ dimensions is not limited to one loop.

An indirect hint of additional cancellations at higher loops comes
from $M$ theory dualities.  In refs.~\cite{StringDuality}, by using
duality (defined in an
appropriate way in the low-energy limit using eleven-dimensional supergravity
with counterterms and a particular string-inspired
regulator), it was argued that in type~II string theory the 
$\partial^4 R^4$ term in the effective action does not suffer 
from renormalization beyond two loops.
Because $D=4,\ \NeqEight$ supergravity may be obtained from
type~II supergravity through compactification of six of the
dimensions, it hints that the former theory may also have additional
cancellations.  A very recent paper from Green, Russo and Vanhove uses
restrictions from $M$ and string theory dualities to argue that $\NeqEight$
supergravity is no worse than logarithmically
divergent in the ultraviolet~\cite{GreenNew}.  
(A related analysis of the four-graviton ten-, twelve- 
and fourteen-derivative terms has also just appeared~\cite{Basu}.)  
Statements linking string dualities to improved ultraviolet
behavior, and perhaps ultraviolet finiteness, in $\NeqEight$ supergravity
may also be found in an earlier paper by Chalmers~\cite{ChalmersDuality}.

An important new clue for the existence of improved ultraviolet
behavior in $\NeqEight$ supergravity comes from the recent work of
Berkovits on the multi-loop effective action of type II string
theories~\cite{Berkovits}. By analyzing string theory amplitudes,
Berkovits found that $\partial^n R^4$ terms in the effective action do
not receive perturbative contributions from above $n/2$ loops for
$0<n<12$, because more than $n+8$ powers of external momenta come out
of the string integrand. 
Here $R^4$ denotes an $\NeqEight$ supersymmetric contraction
of Riemann tensors~\cite{SusyGravity}, and $\partial$ denotes a 
generic spacetime derivative.
Assuming that there are no cancellations
between the massless and higher-mass states of the string
in the loops for small external momenta, the string amplitude properties 
can be applied to supergravity amplitudes, providing an indication 
of additional cancellations in ten-dimensional type~II supergravity.  
If true, then the fact that type~II supergravity corresponds to 
$\NeqEight$ supergravity oxidized to ten dimensions would indicate 
the existence of additional cancellations in four dimensions, 
beyond those of refs.~\cite{BDDPR,HoweStelleNew},
supporting the speculation of Howe and Stelle.  

Finally, the new twistor structure uncovered for gravity
theories~\cite{TwistorGravity,InheritedTwistor} implies a rich set of
constraints on the form of gravity amplitudes.  If $\NeqEight$
supergravity loop amplitudes could be obtained from a topological
string theory, it might lead to a natural explanation for
ultraviolet finiteness.  Recent developments in constructing a
topological twistor string for gravity theories may be found in
ref.~\cite{Mason}.

All these indirect results point to the need to reinvestigate 
the ultraviolet properties of the $\NeqEight$ theory directly.  
A first complete test would be to compute the full
three-loop four-graviton amplitude, in order to confirm that 14 
powers of external momentum do in fact come out of the $\NeqEight$ 
loop momentum integrals.  Here we shall perform a preliminary
examination, studying types of cancellations occurring in three-
and higher-loop amplitudes.

%%%%%%%%%%%%%%%%%%

\section{Unitarity Cuts}

The unitarity method offers a powerful way to determine ultraviolet
properties of gravity theories.  The higher-loop study in
ref.~\cite{BDDPR} relied on using two-particle cuts.  Because of a
remarkable recycling property, it is possible to iterate the
two-particle cuts to {\it all} loop orders.  Although two-particle
iteration provides a wealth of information on the structure of 
the amplitudes, it is only for a limited set of contributions. 
Based on the contributions to the
iterated two-particle cuts, as well as an observed ``squaring''
structure compared to the super-Yang-Mills case, the conclusion of
ref.~\cite{BDDPR} is that the $\NeqEight$ supergravity amplitudes
should be ultraviolet finite for
\begin{equation}
D< {10 \over L} + 2  \hskip 1 cm  (L>1)\,,
\label{OldPowerCount}
\end{equation}
where $D$ is the dimension and $L$ the loop order.  (The case of
one-loop, $L=1$, is special and the amplitudes are ultraviolet finite
for $D<8$, not $D<12$.)  This formula implies that in $D=4$ the first
potential divergence is at five loops.  This result 
was confirmed by studying all cuts, but restricted
to MHV helicity configurations crossing the cuts.  It is also
in agreement with the more recent harmonic superspace analysis of Howe
and Stelle~\cite{HoweStelleNew}. In contrast, the
finiteness condition for $\NeqFour$ super-Yang-Mills theory, 
found in refs.~\cite{BDDPR,HoweStelleNew}, is
\begin{equation}
D < {6 \over L} + 4  \hskip 1 cm (L > 1) \,.
\label{SuperYangMillsPowerCount}
\end{equation}
(For $L=1$, again the amplitudes are finite for $D<8$.)  The bound in
\eqn{SuperYangMillsPowerCount} differs somewhat from the earlier 
superspace power counting bound~\cite{HoweStelleYangMills}, 
though all bounds confirm that $\NeqFour$ 
super-Yang-Mills theory is ultraviolet finite in $D=4$.  In
the planar limit the complete expressions for the $\NeqFour$
super-Yang-Mills amplitudes are known through four
loops~\cite{BRY,BDDPR,ThreeFourLoop}. We have evaluated all
logarithmic singularities of the planar contributions in the critical
dimensions $D_c=7,6,11/2$ corresponding to two, three and four loops,
directly confirming \eqn{SuperYangMillsPowerCount}, at least in the
limit of a large number of colors.  For the $\NeqEight$ supergravity case
there are no complete calculations beyond two loops, so the finiteness
condition (\ref{OldPowerCount}) is much less certain.

%%%%%%%%% FIGURE %%%%%%%%%%%%%%%
\begin{figure}[t]
\centerline{\epsfxsize 2.5 truein \epsfbox{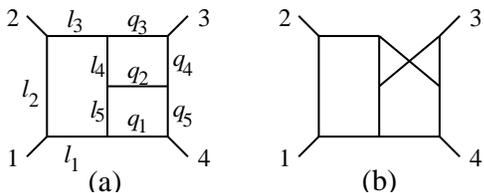}}
\caption[a]{\small Diagram (a) corresponds to a contribution appearing
in the iterated two-particle cut of \fig{CutsFigure}(a). In $\NeqFour$
super-Yang-Mills the iterated two-particle cuts give a numerator
factor of $(l_1+k_4)^2$.  In $\NeqEight$ supergravity it is
$[(l_1+k_4)^2]^2$.
Diagram (b) contains a non-planar contribution which is not
detectable in the iterated two-particle cut of \fig{CutsFigure}(a),
but is detectable in the cut of \fig{CutsFigure}(b).  }
\label{OldPowerFigure}
\end{figure}
%%%%%%%%%%%%%%%%%%%%%%%%%%%%%%%%

A key assumption behind the finiteness condition~(\ref{OldPowerCount}) is
that there are no cancellations with terms not present in iterated
two-particle cuts.  As discussed above, there are good reasons to
reexamine this assumption.

%%%%%%%%% FIGURE %%%%%%%%%%%%%%%
\begin{figure}[t]
\centerline{\epsfxsize 3 truein \epsfbox{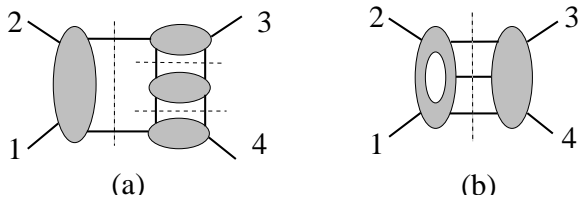}}
\caption[a]{\small  An iterated two-particle cut (a) and a 
three-particle cut (b).}
\label{CutsFigure}
\end{figure}
%%%%%%%%%%%%%%%%%%%%%%%%%%%%%%%%

To do so, consider an $\NeqEight$ supergravity contribution from the
iterated two-particle cuts used in the power count of
ref.~\cite{BDDPR}.  From this reference, the planar contribution,
displayed in \fig{OldPowerFigure}(a), is
\begin{equation}
[s^2 t \, A^\tree_4]^j \int {d^D l_1 \over (2\pi)^D}  {d^D q_1 \over (2\pi)^D}
    {d^D q_2 \over (2\pi)^D} 
   {[(l_1 + k_4)^2]^j \over l_1^2 l_2^2 l_3^2 l_4^2 l_5^2
                q_1^2 q_2^2 q_3^2 q_4^2 q_5^2} \,,
\label{IteratedIntegrand}
\end{equation}
up to overall normalization factors not depending on
momenta. Here $A_4^\tree \equiv A_4^\tree(1,2,3,4)$ is a
color-ordered four-point super-Yang-Mills tree
amplitude, $j=1$ for $\NeqFour$ super-Yang-Mills and $j=2$ for
$\NeqEight$ supergravity.  The $k_i$ are external momenta, labeled by
$i$ in \fig{OldPowerFigure}(a). The Mandelstam
invariants are $s=(k_1+k_2)^2$ and $t = (k_2+k_3)^2$ and the $l_m$ are
the momenta of the left pentagon subintegral in
\fig{OldPowerFigure}(a), while the $q_m$ are the momenta appearing in
the other loops.  The other contributions to the iterated two-particle
cuts are similar.  For the $\NeqFour$ case, an examination of the
three-particle cuts confirms~\cite{BRY,BDDPR} that the correct
numerator factor is $(l_1 + k_4)^2$, instead of, for example $2\, l_1
\cdot k_4$ which is equivalent using the on-shell conditions of the
iterated two-particle cut in \fig{CutsFigure}(a). No analogous check
has been performed on the $\NeqEight$ amplitudes.  For the supergravity
case, we can rewrite the prefactor in front of the integral directly 
in terms of the supergravity tree amplitude, using the KLT-like relation
\begin{equation}
i [s t A_4^\tree(1,2,3,4)]^2 = st (s+t) M_4^\tree(1,2,3,4) \,,
\end{equation}
where $M_4^\tree$ is the $\NeqEight$ four-point tree amplitude.

The scaling of the integral in \eqn{IteratedIntegrand} is that
it is finite for
\begin{equation}
3D < 20 -2j \,,
\end{equation}
so that it corresponds to \eqn{OldPowerCount} 
and \eqn{SuperYangMillsPowerCount} with $L=3$, for $j=2$ and $j=1$ 
respectively.

Is this power counting consistent with the three-particle cut in
\fig{CutsFigure}(b)?  On the left-hand side of the cut we have a
one-loop pentagon integral contribution proportional to
\begin{eqnarray}
\int {d^D l_1 \over (2\pi)^D} \, 
       {[(l_1 + k_4)^2]^j  \over l_1^2 l_2^2 l_3^2 l_4^2 l_5^2}\,.
\label{IteratedTwoParticlePentagon}
\end{eqnarray}
If we perform an integral reduction~\cite{PV}, in the Yang-Mills case,
$j=1$, \eqn{IteratedTwoParticlePentagon} reduces to a sum over box
integrals.  For the $\NeqEight$ supergravity case, $j=2$, a similar
reduction leads also to triangle integrals because of the higher power
of loop momentum in the numerator.  It is important to note that replacing
$[(l_1 + k_4)^2]^2$ with $(2 \, l_1 \cdot k_4)^2$ also leads, under 
integral reduction, to a violation of the no-triangle hypothesis,
although it has a better ultraviolet behavior.

We may compare this power counting
to the results from an evaluation of the three-particle
cuts. On the left-hand side of the cut in \fig{CutsFigure}(b), 
the one-loop five-point supergravity amplitude is given
by~\cite{OneloopMHVGravity}
\begin{eqnarray}
&& \hskip -.7 cm 
M_5^\oneloop(1,2,q_1, q_2,q_3) \label{FiveGravMHV}
 \\
&& \hskip .1 cm \null
= -{1 \over 2} \sum_{\rm perms} s_{q_2 q_1} s_{12}^2 s_{2q_3}^2 
A_5^\tree(1,2,q_3,q_2,q_1)  \nn\\
&& \null \hskip .3 cm 
\times A_5^\tree(1,2,q_3,q_1,q_2) 
\int {d^D l_1 \over (2\pi)^D} {1 \over l_1^2 l_2^2 l_3^2 l_4^2}
 + \Ord(\e) \,, \nn
\end{eqnarray}
where $A_5^\tree$ are color-ordered five-point super-Yang-Mills tree
amplitudes and $s_{2q_3}= (k_2 +q_3)^2$, etc.  In \eqn{FiveGravMHV} we
have dropped the contributions that vanish away from four-dimensions,
since they are suppressed by a power of $\eps = (4-D)/2$ and should be
irrelevant as far as the leading ultraviolet behavior near $D=4$ is
concerned.  At five points all amplitudes are MHV, and there are
simple supersymmetric Ward identities~\cite{SWI} relating the various
five-point amplitudes involving the superpartners; these are described
in appendix E of ref.~\cite{BDDPR}.  The appearance of exactly four
propagators in each term in \eqn{FiveGravMHV} implies that pentagon
integrals have canceled down to boxes, but with no triangle integrals
present.  Alternatively, instead of carrying out the integration one
can use the merging procedure on the cut integrands discussed in
ref.~\cite{TwoLoopSplit} to algebraically arrive at the same
conclusion.  The coefficients of these box integrals may also be
readily obtained using an observation due to Britto, Cachazo and Feng
that the quadruple cuts freeze box integral loop momenta, allowing for
their simple algebraic determination~\cite{BCFUnitarity}.  The lack of
triangle integrals involves a rather non-trivial set of cancellations:
The permutation sum is over 30 contributions corresponding to the
distinct scalar box integrals with one external massive leg.

Comparing \eqn{FiveGravMHV} to \eqn{IteratedTwoParticlePentagon} we
see that the one-loop amplitude entering the cut is much better
behaved in the ultraviolet than is implied by the result
(\ref{IteratedIntegrand}).  Because \eqn{IteratedTwoParticlePentagon}
violates the no-triangle hypothesis, which we know is correct at five
points~\cite{OneloopMHVGravity}, some of the powers of loop momenta
must cancel. However, this cancellation is not visible in the
contribution~(\ref{IteratedIntegrand}).  A crucial difference between
the iterated two-particle cut depicted in \fig{CutsFigure}(a) and the
three-particle cut depicted in \fig{CutsFigure}(b) is that the latter
includes also a variety of other diagrammatic topologies.  For
example, the non-planar diagram (b) of \fig{OldPowerFigure} is not
detected in the iterated two-particle cut of \fig{CutsFigure}(a), 
but it is included in the three-particle cut of \fig{CutsFigure}(b).  
These cancellations might involve integrals that are detectable 
in other iterated two-particle cuts, or integrals that are visible 
only in higher-particle cuts.  In $\NeqFour$ super-Yang-Mills theory, 
cancellations between planar and non-planar topologies cannot happen, 
because planar and non-planar contributions carry different color 
factors and can therefore be treated independently.

%%%%%%%%% FIGURE %%%%%%%%%%%%%%%
\begin{figure}[t]
\centerline{\epsfxsize 2.5 truein \epsfbox{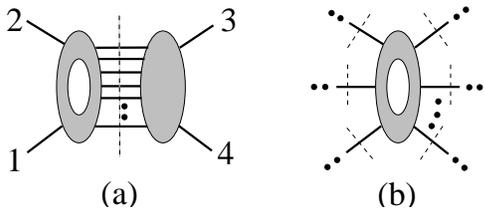}}
\caption[a]{\small From the ``no-triangle hypothesis'' all one-loop
subamplitudes appearing in unitarity cuts in $\NeqEight$
supergravity have the same degree of divergence as in $\NeqFour$
super-Yang-Mills theory.  The cut (a) is an $L$-particle cut of
an $L$-loop amplitude. Cut (b) makes use of generalized unitarity;
if a leg is external to the entire amplitude, it should not be cut.}
\label{NCutFigure}
\end{figure}
%%%%%%%%%%%%%%%%%%%%%%%%%%%%%%%%

\begin{figure}[t]
\centerline{\epsfxsize 1. truein \epsfbox{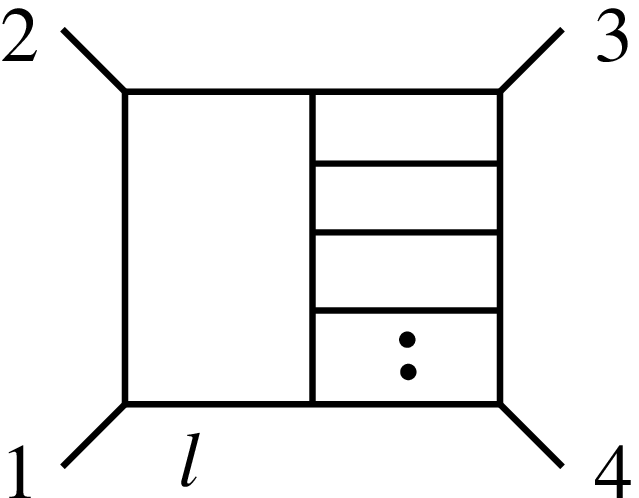}}
\caption[a]{\small An $L$-loop contribution as proposed in
ref.~\cite{BDDPR}. In $\NeqFour$ super-Yang-Mills theory the 
numerator factor is $[(l+ \nobreak k_4)^2]^{(L-2)}$, 
while in $\NeqEight$ supergravity the factor is
$[(l+k_4)^2]^{2(L-2)}$.}
\label{NOldPowerFigure}
\end{figure}

Cancellations at higher loops are also dictated by the no-triangle
hypothesis.  Consider the $L$-particle cut of the $L$-loop amplitudes
shown in \fig{NCutFigure}(a). The iterated two-particle cut analysis
and squaring assumption of ref.~\cite{BDDPR} give the numerator
factor appearing in the diagrammatic contribution shown in
\fig{NOldPowerFigure} as $[(l + k_1)^2]^{2(L-2)}$ for the $\NeqEight$
case.  This factor is the square of the $\NeqFour$ super-Yang-Mills factor
$[(l + k_1)^2]^{(L-2)}$.  This proposed $\NeqEight$ numerator factor
is at the origin of the power count~(\ref{OldPowerCount}).  However,
this factor leads to a violation of the no-triangle hypothesis for the
$(L+2)$-leg amplitude on the left-hand side of the $L$-particle cut in
\fig{NCutFigure}(a). That is, we find that the cancellations which
reduce the one-loop amplitude to a sum over box integrals have not
been taken into account.  Since there is strong evidence in favor of
the one-loop no-triangle hypothesis~\cite{NoTriangle}, we conclude
that the finiteness condition (\ref{OldPowerCount}) is probably too
conservative.

%%%%%%%%% FIGURE %%%%%%%%%%%%%%%
%%%%%%%%%%%%%%%%%%%%%%%%%%%%%%%%

One can also extend this analysis using generalized unitarity, which
provides a powerful way to construct
amplitudes~\cite{GeneralizedUnitarity, TwoLoopSplit,
BCFUnitarity}. For all possible one-loop sub-amplitudes isolated by
cutting internal lines in a higher-loop amplitude, as depicted in
\fig{NCutFigure}(b), the no-triangle hypothesis implies that they have
the same degree of divergence as the $\NeqFour$ super-Yang-Mills
theory.  Because this result holds for all possible generalized cuts which
isolate a one-loop amplitude, we obtain a rather non-trivial set of
consistency conditions limiting the ultraviolet behavior of the
higher-loop amplitudes.  If a specific set of cuts points to bad
ultraviolet behavior in a given loop momentum, we can isolate that
loop via generalized unitarity.  Every such one-loop
subamplitude has a power count no worse than that of $\NeqFour$
super-Yang-Mills theory (assuming the no-triangle hypothesis), 
suggesting that the entire amplitude may have this property. 

%%%%%%%%% FIGURE %%%%%%%%%%%%%%%
\begin{figure}[t]
\centerline{\epsfxsize 1.1 truein \epsfbox{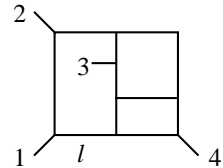}}
\caption[a]{\small An example of a potential $\NeqEight$ supergravity
contribution where the no-triangle hypothesis does not provide
sufficient information to rule out a violation of the 
bound (\ref{SuperYangMillsPowerCount}).  The numerator factor
proposed in ref.~\cite{BDDPR} is $[(l+k_4)^2]^2$, although 
when evaluated on the iterated two-particle cuts used in its construction, 
it is indistinguishable from $(2\, l\cdot k_4)^2$, which is consistent
with the no-triangle bound.}
\label{BadPowerFigure}
\end{figure}
%%%%%%%%%%%%%%%%%%%%%%%%%%%%%%%%

In order to construct a proof that the overall degree of divergence
matches that of $\NeqFour$ super-Yang-Mills, it is crucial to track the
critical dimension where logarithmic divergences first arise.  In such
a proof one would need to rule out contributions where the no-triangle
hypothesis is not violated, yet the overall finiteness bound
(\ref{SuperYangMillsPowerCount}) is violated. An example of such a
potential contribution is given in \fig{BadPowerFigure}. The numerator
factor proposed in ref.~\cite{BDDPR}, $[(l+k_4)^2]^2$, would not violate
the no-triangle hypothesis for the one-loop hexagon subdiagram, yet
would violate the overall bound.  On the other hand, the iterated
two-particle cut analysis used in its construction~\cite{BDDPR} does
not distinguish $[(l+k_4)^2]^2$ from $(2\, l \cdot k_4)^2$
--- the latter form {\it is} consistent with the overall
bound~(\ref{SuperYangMillsPowerCount}) 
--- nor does it take into account any cancellations of the type 
found for the planar contribution in \fig{OldPowerFigure}.  
A complete construction of the three-loop amplitude would, 
of course, resolve this situation.

By power counting Feynman diagrams, one can see that if the finiteness
condition of $\NeqEight$ supergravity is identical to that of
$\NeqFour$ super-Yang-Mills theory, then the $L$-loop contribution to
the one-particle irreducible effective action would start with the
form $\partial^{2L} R^4/\partial^6$, with an ultraviolet-finite
(though infrared-singular) coefficient in $D=4$.  The nonlocal factor
of $1/\partial^6$ arises from the loop integrals, by dimensional
analysis.  A discussion of power counting in effective actions, and
implications for the degree of divergence of the theory in $D=4$ may
be found in ref.~\cite{GreenNew}.  
The precise derivative factors that actually
appear would need to be calculated.  At two loops, the explicit
integrand for the amplitude~\cite{BDDPR}, as well as the values of the
dimensionally regularized infrared singular integrals in
$D=4$~\cite{DoubleBoxIntegrals}, are known.

%%%%%%%%%%%%%%%%%%%%%%%%%%%

\section{Discussion}

In this note we discussed evidence that four-dimensional $\NeqEight$
supergravity may be ultraviolet finite.  Given the additional
cancellations we observe at higher loops,
as well as the other clues described in the 
introduction~\cite{ChalmersDuality, HoweStelleNew, StringDuality, 
InheritedTwistor, SixPtIRGravity, Berkovits, NoTriangle, GreenNew}, 
there is good reason to believe that the finiteness bound of
refs.~\cite{BDDPR,HoweStelleNew} is too conservative.  Clearly a
closer direct reexamination of the ultraviolet properties of
$\NeqEight$ supergravity is needed.  A number of calculations should
be carried out to this end. The most important task is to construct
complete amplitudes beyond two loops. One could then check directly
whether they satisfy the same $D$-dimensional finiteness
condition~(\ref{SuperYangMillsPowerCount}) obeyed by $\NeqFour$
super-Yang-Mills amplitudes.

Using the unitarity method it is feasible to construct
complete four-point integrands at three and perhaps higher loops.
As explained in ref.~\cite{BDDPR}, higher-loop calculations 
of the $\NeqEight$ supergravity unitarity cuts are essentially
double copies of $\NeqFour$ super-Yang-Mills cuts, due to
the KLT relations. In the super-Yang-Mills case, the complete planar
four-point three- and four-loop integrands are
known~\cite{BRY,ThreeFourLoop}. For supergravity, one also needs 
non-planar contributions, which are more complicated than planar ones.  

In order to confirm that the cancellations are not limited to
divergences of the form $\partial^n R^4$, but extend to operators with
more powers of $R$, it is important to construct integrands for
higher-point amplitudes.  Given that the five-point two-loop planar
$\NeqFour$ super-Yang-Mills integrand has already been
determined~\cite{TwoLoopFivePoint}, it should also be feasible to
obtain the five-point two-loop $\NeqEight$ supergravity amplitude. 

It should also be possible to carry out all-order studies using the
unitarity method, given the recursive nature of the formalism.
Tracking potential logarithmic divergences that arise in the critical 
dimension $D_c$ is crucial.  Such divergences are unambiguous, 
whereas power divergences can depend on details of the regularization scheme.

Although there is already rather strong evidence that in
$D$ dimensions one-loop $\NeqEight$ supergravity amplitudes have the
same degree of divergence as their $\NeqFour$ super-Yang-Mills
counterparts~\cite{OneloopMHVGravity, InheritedTwistor,
SixPtIRGravity, NoTriangle}, it is important to construct a
complete proof, because the result is a key ingredient for 
using the unitarity method in higher-loop analyses.

At the multi-loop level, besides carrying out explicit constructions
of complete amplitudes, it would also be important to identify an underlying
dynamical principle or symmetry explaining the additional cancellations
observed.

\section*{Acknowledgments}
We thank Nathan Berkovits, Emil Bjerrum-Bohr, John Joseph Carrasco,
Dave Dunbar, Harald Ita, Henrik Johansson, David Kosower and Arkady
Tseytlin for helpful discussions.  This research was supported by the
US Department of Energy under contracts DE--FG03--91ER40662 and
DE--AC02--76SF00515, and the National Science Foundation under grant
PHY-0608114.

%%%%%%%%%%%%%%%%%%%%%%%%%%%%%%%%%%%%%%%%%%%%%%%%%%%%%%%%%%%

\end{document}